\documentclass[aps,pra,twocolumn,groupedaddress,superscriptaddress]{revtex4-1}
\usepackage{bm}
\usepackage[pdftex]{graphicx}
\usepackage{amsmath}
\begin{document}

%\preprint{}

%\title{Geometrically varying phase matching condition and time evolution of coherent two-photon emission}
%\title{Angular dependent phase matching condition and time evolution of nearly degenerate four-wave mixing using two-photon transition of hydrogen molecules}
%\title{Nearly degenerate four-wave mixing using coherently amplified two-photon transition of hydrogen molecules}
%\title{Frequency and time dependence of two-photon resonant nearly degenerate four-wave mixing}
\title{Geometry-dependent spectra and coherent-transient measurement \\
of nearly degenerate four-wave mixing using two-photon resonance}
%\\
%from Hydrogen Molecules}


%%\author{Hideaki Hara, 
%$^1$
%%Yuki~Miyamoto, Takahiro~Hiraki, Takahiko~Masuda, Noboru~Sasao, Satoshi~Uetake, 
%%Akihiro~Yoshimi, Koji~Yoshimura, and Motohiko~Yoshimura}
%\email[1]{hhara@okayama-u.ac.jp
%\thanks{}
%%\affiliation{Research Institute for Interdisciplinary Science, Okayama University, Okayama 700-8530, Japan}

\author{Hideaki~Hara}
\altaffiliation{hhara@okayama-u.ac.jp}
\affiliation{Research Institute for Interdisciplinary Science, Okayama University, Okayama 700-8530, Japan}
\author{Yuki~Miyamoto}
\affiliation{Research Institute for Interdisciplinary Science, Okayama University, Okayama 700-8530, Japan}
\author{Takahiro~Hiraki}
\affiliation{Research Institute for Interdisciplinary Science, Okayama University, Okayama 700-8530, Japan}
\author{Kei~Imamura}
\affiliation{Research Institute for Interdisciplinary Science, Okayama University, Okayama 700-8530, Japan}
\author{Takahiko~Masuda}
\affiliation{Research Institute for Interdisciplinary Science, Okayama University, Okayama 700-8530, Japan}
\author{Noboru~Sasao}
\affiliation{Research Institute for Interdisciplinary Science, Okayama University, Okayama 700-8530, Japan}
\author{Satoshi~Uetake}
\affiliation{Research Institute for Interdisciplinary Science, Okayama University, Okayama 700-8530, Japan}
\affiliation{PRESTO, Japan Science and Technology, Okayama 700-8530, Japan}
\author{Akihiro~Yoshimi}
\affiliation{Research Institute for Interdisciplinary Science, Okayama University, Okayama 700-8530, Japan}
\author{Koji~Yoshimura}
\affiliation{Research Institute for Interdisciplinary Science, Okayama University, Okayama 700-8530, Japan}
\author{Motohiko~Yoshimura}
\affiliation{Research Institute for Interdisciplinary Science, Okayama University, Okayama 700-8530, Japan}


\date{\today}

\begin{abstract}
We study nearly degenerate four-wave mixing 
using a two-photon-allowed vibrational transition of parahydrogen.
A signal photon is generated by a trigger photon and coherence among parahydrogen, 
which is prepared by two counterpropagating pump pulses.
The dependence of the signal pulse energy on the trigger frequency are investigated.
The measured spectra vary depending on the geometry.
They shift depending on the direction of the signal pulse 
and on the small angle formed by the counterpropagating pump pulses.
Furthermore, the dependence of signal pulse energy 
on the incident time of the trigger pulse is investigated.
The measured signal pulse energy is high 
if the trigger pulse is slightly delayed with respect to the pump pulses.
We demonstrate that 
these geometry-dependent spectra and coherent-transient response can be explained 
by using simple models.
\end{abstract}

%\pacs{42.50.Gy,32.80.Qk, 33.80.-b, 42.62.Lm}
%\keywords{}

\maketitle

%%\section{}
%%\subsection{}
%%\subsubsection{}

\section{Introduction}
\label{sec:introduction}

Four-wave mixing is a third order nonlinear optical phenomenon \cite{Boyd-text-2008} 
and has various applications such as optical parametric generation in the deep-ultraviolet 
\cite{FWM-UVgeneration-1997,FWM-UVgeneration-2007}, 
supercontinuum generation \cite{SC-1970,SC-review-2006}, 
and coherent anti-Stokes Raman scattering spectroscopy 
\cite{CARS-1965,CARS-1974}.
If the frequencies of four photons are close to each other, 
the process is referred to as nearly degenerate four-wave mixing (NDFWM).
This process has been often investigated on the situations 
where two intense counterpropagating pump fields with the same frequency $\omega$ 
and a third weak trigger field with a slightly different frequency $\omega - \Delta \omega$ 
are injected.
If phase mismatch is not too large, 
a fourth signal field with a frequency of $\omega + \Delta \omega$ 
developes constructively 
and propagates along the direction opposite to that of the third trigger field.
This process can yield an optical narrow bandpass retroreflector 
whose bandwidth is much narrower than those of commercial bandpass filters 
\cite{bandpass-Pepper-1978,bandpass-Nilsen-1979}.

The use of resonances between actual levels enhances nonlinear susceptibility 
in four-wave mixing processes.
Coherence among ensemble of atoms or molecules is generated 
and a high signal intensity is provided within coherence time in this case.

Resonant NDFWM using counterpropagating pump fields has been studied so far.
For the case using a one-photon transition, 
the dependence on the frequency detuning $\Delta \omega$ has been studied 
theoretically \cite{bandpass-Pepper-1978,bandpass-Nilsen-1979,NDFWM-ACStark-1980}  and experimentally \cite{bandpass-Nilsen-exp-1981,NDFWM-ACStark-1981}.
The measured spectrum in Ref. \cite{bandpass-Nilsen-exp-1981} 
had a peak at $\Delta \omega = 0$ 
and a bandwidth much narrower than that determined by the phase matching condition.
For the case using a two-photon transition, 
it has been investigated only theoretically \cite{two-photon-NDFWM-1981} 
and no experimental data have been reported as yet.
Agrawal predicted the frequency spectrum of the fourth signal field 
using one-dimensional coupling equations between the third trigger and fourth signal fields 
\cite{two-photon-NDFWM-1981}.
Applying the obtained result to the weak coupling limit 
between the trigger and signal fields, 
the spectrum is expected to be a square of a sinc function 
whose peak frequency detuning is $\Delta \omega = 0$.

The time dependence of resonant degenerate four-wave mixing has also been investigated.
For two-photon resonance cases, 
decay of signal intensity 
was monitored with changing the incident time of the third trigger field 
\cite{time-resolved-two-photon-DFWM-1977,time-resolved-two-photon-DFWM-1979}.
These previous studies focused on 
only the time region in which the signal intensity decayed 
after the pump pulses passed away, 
but they did not investigate in detail the time region 
in which the incident pump fields still existed in the target.
The signal pulse intensity is higher in this time region than in the decay time region.
It is important to know the time when the signal intensity becomes the maximum 
to apply NDFWM as a bandpass retroreflector.
It would be determined by the both effects of coherence development and decoherence.

\begin{figure}[t]
\centering
\includegraphics[width=4cm,keepaspectratio,pagebox=artbox]{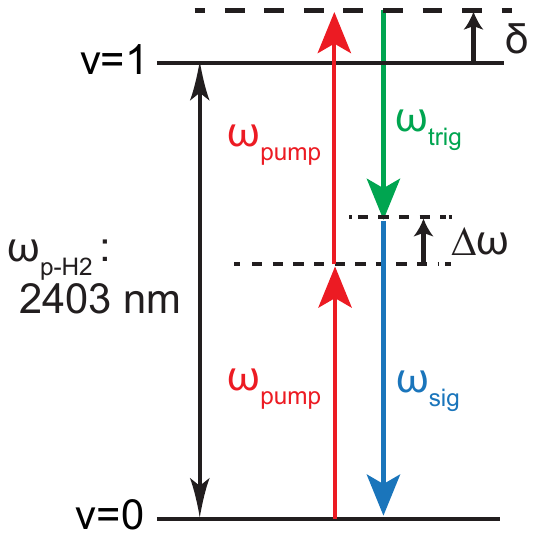}[]
\caption{(Color online) 
Energy diagram of p-H$_{2}$.
The energy difference between the $v = 0$ and $v = 1$ vibrational states 
corresponds to a wavelength of 2403~nm 
($\omega_{\mathrm{p \mathchar`- H_{2}}}$).
The frequencies $\omega_{\mathrm{pump}}$, $\omega_{\mathrm{trig}}$, 
and $\omega_{\mathrm{sig}}$ are those of 
the pump, trigger for TPE, and TPE signal, respectively.
The two-photon detuning $\delta$ can be shifted 
by changing the wavelength of the pump pulses.
$\Delta \omega$ is the frequency detuning of the TPE signal 
from $\omega_{\mathrm{pump}}$.}
\label{fig:energydiagram}
\end{figure}

Recently, we succeeded in degenerate four-wave mixing using two-photon resonance 
of the vibrational transition of 
gaseous parahydrogen molecules (p-H$_{2}$) \cite{taiko-2018}.
We observed two-photon emission (TPE) signal 
which is generated as a result of degenerate four-wave mixing.
The energy diagram is shown in Fig. \ref{fig:energydiagram}.
All the four interacting waves had the identical frequency 
($\omega_{\mathrm{pump}} = \omega_{\mathrm{trig}} = \omega_{\mathrm{sig}}$).
The p-H$_{2}$ molecules are excited 
by two-photon absorption of counterpropagating photons.
This transition in p-H$_{2}$ is suitable for observation of TPE process.
It is because single-photon $E1$ transitions between vibrational levels of 
homonuclear molecules are forbidden in the selection rules, 
but two-photon $E1 \times E1$ transitions are allowed.
The spontaneous TPE rate for p-H$_{2}$ is $\mathcal{O} (10^{-11})$~Hz, 
but coherence generated by the excitation lasers enhance the transition rate 
like superradiance \cite{SR-1954}.
We also measured dependences of the coherently amplified TPE signal energy on 
the detuning $\delta$, target pressure, and input pulse energies, 
and demonstrated that the measured results were 
qualitatively consistent with a simulation using the Maxwell-Bloch equations.

In the present paper, 
we vary frequency and incident time of the trigger pulses 
independently of those of pump pulses.
Applying the usual one-dimensional coupling equations \cite{two-photon-NDFWM-1981} 
to our experimental condition in the weak coupling limit, 
the frequency spectrum of the TPE signal field 
is expected to be a square of a sinc function 
whose center frequency detuning is $\Delta \omega = 0$ 
and width (FWHM) is roughly 1~GHz.
The width depends on the interaction length of the pump and trigger fields, 
but the center frequency detuning does not change depending on it.
This spectrum is calculated under the assumption 
that the signal field propagates on the same axis as the trigger field 
and that the pump fields propagate completely in the opposite direction to each other.
The measured spectra are different from the expected sinc-squared function 
and have the peaks at $\Delta \omega \neq 0$ and the widths broader than 1~GHz.
We demonstrate the measured spectra by a simple model excluding the above assumptions.
We also show the TPE signal intensity is high 
if the trigger pulse is slightly delayed with respect to the pump pulses.
It means that the TPE signal is not just temporal overlap 
of the incident pulses, it would reflect the time evolution of coherence among p-H$_{2}$.
We demonstrate that 
this coherent-transient response 
can be explained by another simple model using the optical Bloch equations.

\section{Experimental Setup}
\label{sec:setup}

\begin{figure}[t]
\centering
\includegraphics[width=8cm,keepaspectratio,pagebox=artbox]{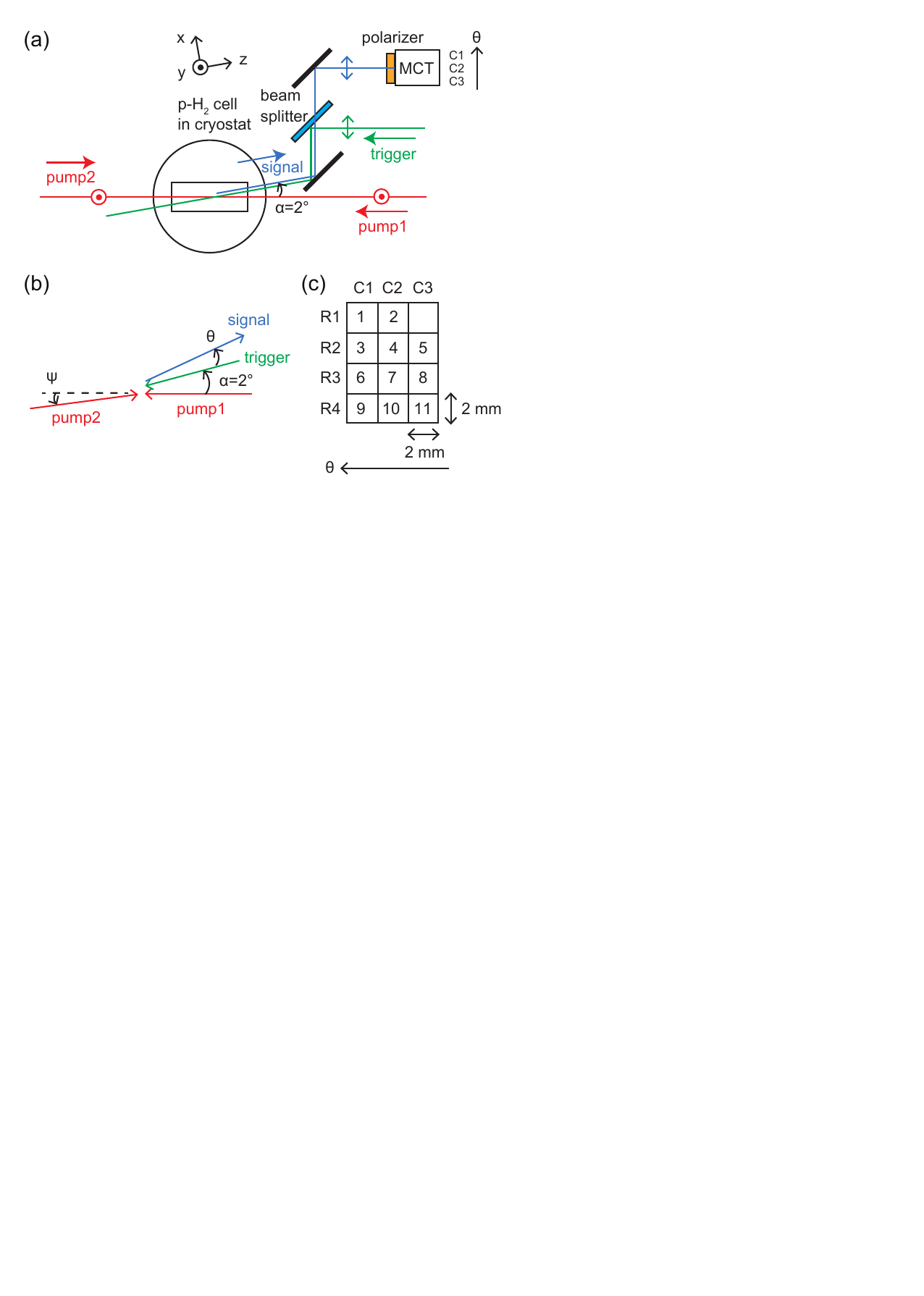}
\caption{(Color online) 
(a) Experimental setup.
The red, green, and blue lines are the paths of pump, trigger, and TPE signal fields.
The leftwards and rightwards arrows represent the propagation directions.
The circles and up down arrows represent the vertical and horizontal polarizations.
$z$ axis is defined as the opposite to the propagation direction of the trigger field.
$x$, $y$, and $z$ axes are defined so that they construct a right-handed system.
The coordinate origin is at the center of the target.
(b) Angular relationship between the pump1, pump2, trigger, and TPE signal fields.
The actual angles $\alpha$, $\theta$, and $\psi$ are very small, 
but they are exaggerated.
(c) Position of the detector.
The eleven numbered regions are used for measurement of spectra.
The rows and columns are labeled as R1--4 and C1--3, respectively.
The arrow represents the $\theta$ positive direction.
}
\label{fig:setup}
\end{figure}

Figure \ref{fig:setup} (a) shows 
the experimental setup around the target.
The counterpropagating pump fields, 
which are referred to as pump1 and pump2, 
are injected into the p-H$_{2}$ target cell from the both sides.
If coherence is prepared by the counterpropagating pump fields, 
the trigger photons induce the coherently amplified TPE process.
It generates pairs of another trigger photon and a TPE signal photon.
The frequencies of the emitted two photons satisfy energy conservation 
($\omega_{\mathrm{trig}} + \omega_{\mathrm{sig}} = 2 \omega_{\mathrm{pump}}$).
They are denoted as 
$\omega_{\mathrm{trig}} = \omega_{\mathrm{pump}} - \Delta \omega$  
and $\omega_{\mathrm{sig}} = \omega_{\mathrm{pump}} + \Delta \omega$, 
by using a frequency detuning $\Delta \omega$.
We change $\Delta \omega$ by tuning the trigger frequency.
The mutual incident timing between the pump and trigger fields can be adjusted.
The TPE signal field propagates in the direction opposite to the trigger field 
due to the phase matching condition.
The trigger field is injected at an angle $\alpha = 2^{\circ}$ to the pump fields 
to separate the TPE signal from the pump fields.
The slight angle is necessary for the long interaction length.
The TPE signal field is separated from the path of the trigger field 
by using the beam splitter (Thorlabs, BSW511).
The TPE signal energy is measured 
by a mercury--cadmium--telluride (MCT) mid-infrared detector (Vigo systems, PC-3TE-9).

Figure \ref{fig:setup} (b) shows 
the angular relationship between the pump, trigger, and TPE signal fields.
These angles are not limited in the horizontal plane, but three dimensional.
The angle between the pump fields is denoted as $\psi$.
The angle between the trigger and TPE signal fields is denoted as $\theta$.
In the present paper we shift the position of the MCT detector, 
whose size of acceptance surface is 2~mm $\times$ 2~mm, 
horizontally and vertically 2~mm by 2~mm 
and measure the spectrum at each position.
The distance from the center of the target to the detector is roughly 500~mm.
The size of acceptance surface of the MCT detector corresponds to the angle range of 
$\Delta \theta \times \Delta \theta = 0.2^{\circ} \times 0.2^{\circ}$ 
and the solid angle of $2 \times 10^{-5}$~sr.
The TPE signal can be detected in the range of roughly 6~mm $\times$ 8~mm.
The eleven regions numbered in Fig. \ref{fig:setup} (c) 
are used for measurement of spectra.
The TPE signal intensity is too small to observe 
in the outer side of these eleven regions.
No lenses are used between the target and the detector 
because the information on the angle $\theta$ would be lost.

We perform similar measurements changing the angle $\psi$.
The angle $\psi$ is changed by the alignment of the pump2 field.
The angle $\psi$ itself is difficult to be measured, 
but the shift of the angle is estimated by measuring the positions of the pump2 field 
in front of and behind the target with a beam profiler.

The p-H$_{2}$ gas is enclosed within a copper cell 
with a diameter of 20~mm and a length of 150~mm.
It is prepared by converting normal hydrogen gas 
in a magnetic catalyst Fe(OH)O that is cooled to roughly 14~K.
The temperature of the cell is kept at 78~K by liquid nitrogen in a cryostat, 
at which temperature almost all the p-H$_{2}$ molecules exist in the ground state.
The pressure of the p-H$_{2}$ gas is 260~kPa, 
which corresponds to a density of $2.4 \times 10^{20}$~cm$^{-3}$.
Both sides of the cell and the cryostat are 
sealed with anti-reflection coated barium fluoride (BaF$_{2}$) windows 
(Thorlabs, WG01050-E).

The two counterpropagating pump fields for coherence generation 
have the identical wavelength of $\lambda = 4806$~nm.
Lasers with high intensity and narrow linewidth are required 
in order to generate high coherence.
The pump pulses are generated 
by using injection-seeding different frequency generation.
The repetition rate and the duration (FWHM) are 10~Hz and 4~ns, respectively.
The linewidth (FWHM) is roughly 150~MHz, 
which is measured by the absorption spectroscopy of 
the rovibrational transition of the carbonyl sulfide gas.
The detailed laser setup is the same as that in Ref. \cite{pH2-THG-2017,taiko-2018}.
The pump pulses are divided by a beam splitter into the pump1 and pump2 pulses.
They are nearly collimated at the target cell and the waist sizes are $w_{0}=$ 1.1~mm.
The input pulse energies of both pump pulses are roughly 1.5~mJ/pulse.

\begin{figure}[t]
\centering
\includegraphics[width=7.5cm,keepaspectratio,pagebox=artbox]{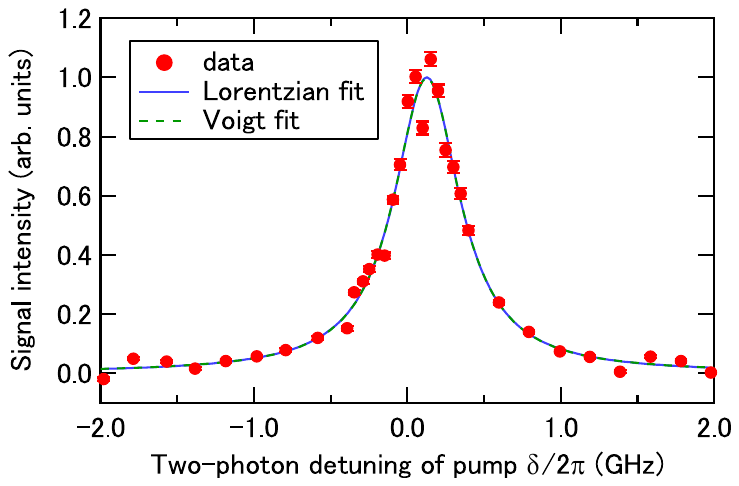}
\caption{(Color online) 
Spectra of the TPE signal as a function of the two-photon detuning $\delta / 2 \pi$.
The red circles are the measured data.
The error bars indicate standard errors.
The blue solid and green dashed lines represent the Lorentzian and Voigt fits to the data, 
respectively.
The result curve of the Voigt fit is almost the same as the Lorentzian fit.
}
\label{fig:pump}
\end{figure}

Two-photon detuning $\delta$ from the resonance, defined as 
$\delta = 2 \omega_{\mathrm{pump}}-\omega_{\mathrm{p \mathchar`- H_{2}}}$, 
is adjusted to maximize the TPE signal energy.
It can be tuned within several GHz.
Figure \ref{fig:pump} shows the spectrum of the TPE signal 
as a function of $\delta / 2 \pi$.
The trigger frequency is fixed at roughly half of 
$\omega_{\mathrm{p \mathchar`- H_{2}}}$.
Two hundred data are taken at each detuning.
The blue line represents the Lorentzian fit.
The center of the Lorentzian is $\delta / 2 \pi = +0.13$~GHz.
The detuning uncertainty is $\pm0.17$~GHz, 
which is estimated from the absolute accuracy of the wave meter (HighFinesse, WS-7). 
The TPE signal 
is maximum at resonance within the uncertainty.
We treat $\delta$ as zero in the present paper.
The width (FHWM) of the Lorentzian is $0.52 \pm 0.01$~GHz.
One of the origins of the width is the pressure broadening.
It is estimated to be 0.39~GHz \cite{pH2-Raman-1986}.
The Doppler shifts due to the counterpropagating pump fields are canceled out 
in the present experiment.
The residual component of the width would come from the laser linewidth and others.
We also fit the Voigt function to the data 
taking into account that the frequency distribution of the laser pulse is close to Gaussian.
The result curve of the Voigt fit is almost the same as the Lorentzian fit.
The center of the Voigt fit is $\delta / 2 \pi = +0.13$~GHz.
The FWHMs of the Lorentzian and Gaussian components are 
$0.52 \pm 0.14$~GHz and $0.06 \pm 0.08$~GHz, respectively.
The width of the Gaussian component 
is consistent with zero within the uncertainty.
It indicates that the Doppler shifts are negligibly small.
On the other hand, 
the width of Gaussian component is also consistent with the laser linewidth.

The wavelength of the trigger field is also 4806~nm.
The trigger pulses are also generated 
by using injection-seeding different frequency generation.
The repetition rate and the duration of the trigger pulses are 10~Hz and 2~ns, respectively.
The measured linewidth (FWHM) is roughly 1 GHz.
The frequency of the trigger pulses is tuned within 15~GHz in the present experiment.
The detailed laser setup is described in Ref. \cite{MIR-2017}.
The trigger pulses are loosely focused at the center of the target cell 
and the waist size is $w_{0}=$ 0.4~mm.
The input pulse energy is roughly 5~$\mu$J/pulse, 
which is much weaker than those of the pump pulses.
Coherence is generated mainly by the pump pulses rather than the trigger pulses.

Besides the TPE process third-order harmonic generation (THG) processes occur 
due to the two-photon absorption 
with only the pump1 pulses or the pump2 pulses.
In the former case 
the generated THG photons do not disturb the measurement of the TPE signal 
because they propagate in the opposite direction to the MCT detector.
In the latter case 
a portion of the generated THG photons may reach the MCT detector, 
but they are cut off by a germanium base neutral density filter (Edmund, $\#64-355$).

The scatterings of the incident photons are causes of background.
The scattering photons due to the pump pulses, 
whose polarization is designed to be orthogonal to that of the signal photon, 
can be cut off by the polarizer.
On the other hand, 
the scattering photons due to the trigger pulses is difficult to be separated.
It is the main component of the background.
The offset of the spectrum due to the stray photons 
is subtracted from the measured data in the present paper.

Shot-by-shot fluctuations of pulse energy 
and timing difference between pump and trigger fields are causes of errors.
Energy and timing of each pulse are monitored 
by measuring a small portion of the pulse 
using an indium-antimonide detector (Hamamatsu, P5968-100) for the pump pulses 
and another MCT detector (Vigo systems, PEM-10.6) for the trigger pulses.
The relative standard deviations of each pulse energy of pump and trigger fields 
are roughly $8 \%$ and $20 \%$, respectively.
The standard deviation of timing difference 
between pump and trigger pulses is roughly 0.8~ns.
Four hundred data are taken at each trigger frequency and detector position 
for the measurement of spectra of TPE signal in Sec. \ref{subsec:result-frequency}.
Only the data taken under the condition 
where pump and trigger pulse energy and timing jitter are within $\pm 1.0 \sigma$ 
are used for analysis in order to reduce the errors.
Only the data taken under the condition 
where both pump and trigger pulse energies are within $\pm 0.5 \sigma$ 
are used for analysis of the coherent-transient measurement of TPE signal 
in Sec. \ref{subsec:result-timing}.

\section{RESULTS}
\label{sec:result}

\subsection{Spectra of TPE signal}
\label{subsec:result-frequency}

\begin{figure}[t]
\centering
\includegraphics[width=8.5cm,keepaspectratio,pagebox=artbox]{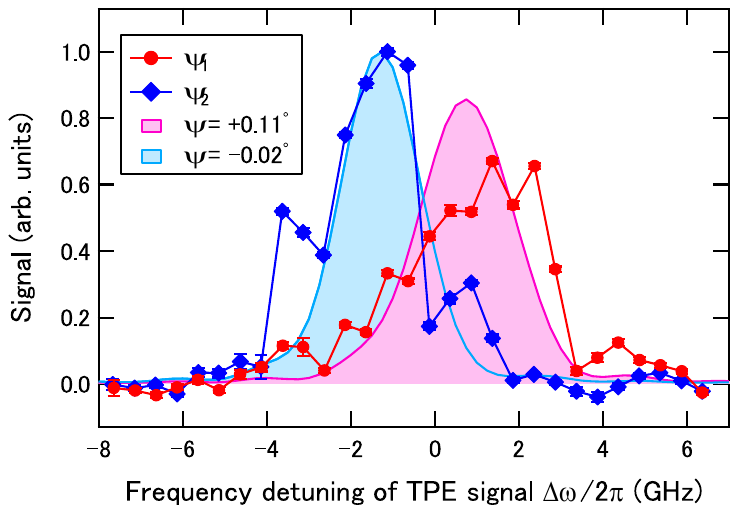}
\caption{(Color online) 
Measured and calculated spectra 
obtained by the sums of TPE signal intensity at each position 
as a function of the frequency detuning of TPE signal $\Delta \omega / 2 \pi$.
The circles and diamonds represent the measured spectra.
The error bars of the measured data indicate standard errors.
The shaded areas represent the calculated spectra.
The spectra are normalized 
so that the maximum values of TPE signal 
for the angle $\psi_{2}$ (experiment) and $\psi = -0.02^{\circ}$ (calculation) 
are equal to unity.
}
\label{fig:total}
\end{figure}

We investigate the dependence of TPE signal pulse energy 
on the frequency detuning $\Delta \omega$.
Figure \ref{fig:total} shows the spectra 
obtained by the sums of the TPE signal 
at each detector position shown in Fig. \ref{fig:setup} (c).
They correspond to the spectra of total TPE signals.
The red circles and blue diamonds represent the measured data 
for the angle $\psi$ of $\psi_{1}$ and $\psi_{2}$, respectively.
The difference of the angle measured with a beam profiler 
is $\Delta \psi_{\mathrm{meas}} = \psi_{2} - \psi_{1} = - 0.13^{\circ}$.
The shapes of the measured spectra are different from the expected sinc-squared function.
The spectrum for $\psi_{1}$ is not symmetric with respect to a peak 
and has a tail in the lower frequency detuning regions.
The peak frequency detunings are 
$+1.4$~GHz for $\psi_{1}$ and $-1.1$~GHz for $\psi_{2}$, respectively.
They are not zero even though the uncertainty of the wave meter is taken into account.
The spectra shift depending on the angle $\psi$.
The peak frequency detuning for $\psi_{2}$ is roughly 2.5~GHz lower 
than that for $\psi_{1}$.
The FWHMs of the measured spectra are 
broader than that of the expected sinc-squared function (1~GHz).
The signal intensity for $\psi_{2}$ is higher than that for $\psi_{1}$.

\begin{figure*}[t]
\centering
\includegraphics[width=18cm,keepaspectratio,pagebox=artbox]{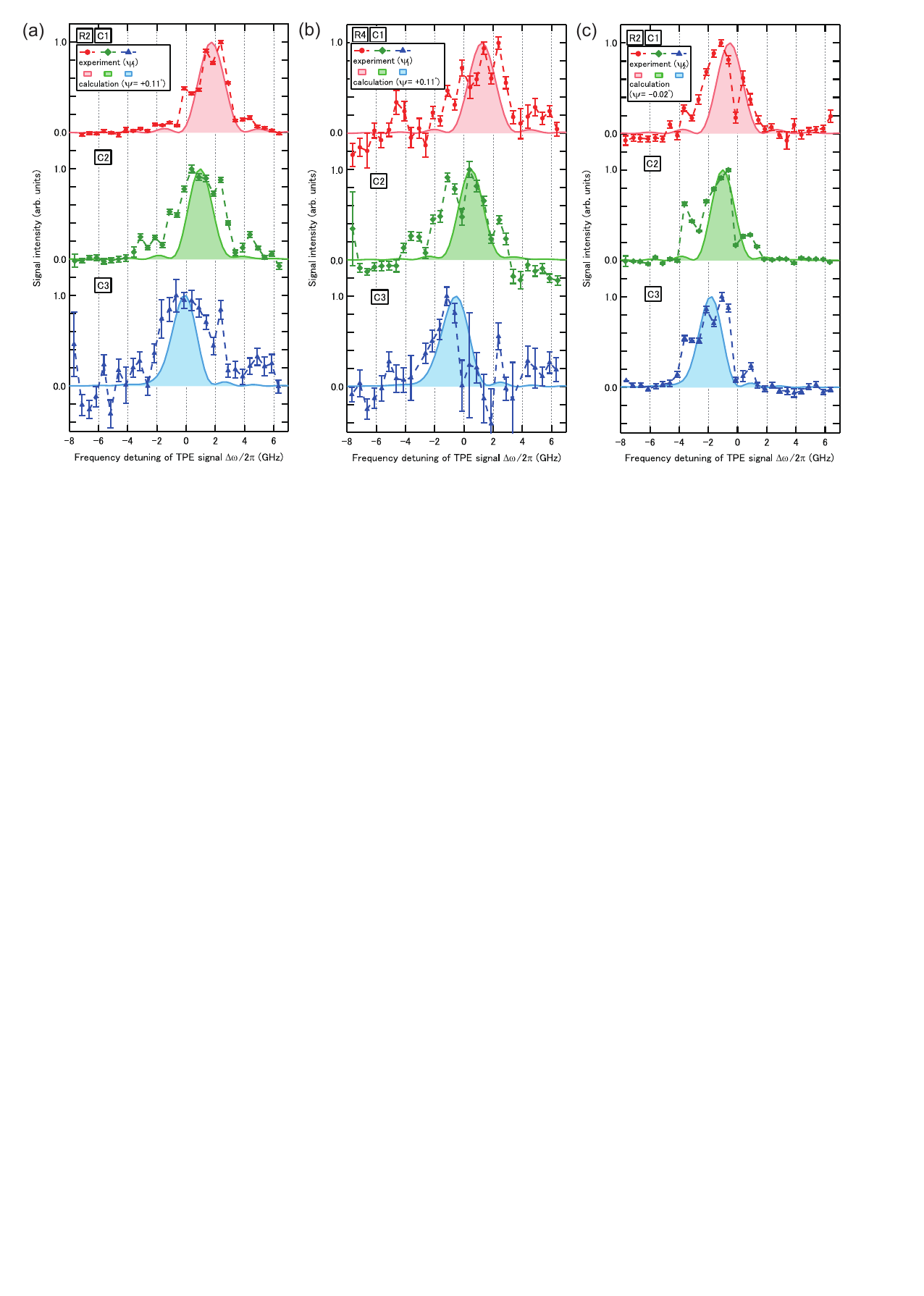}
\caption{(Color online) 
Examples of measured and calculated spectra at each detector position.
(a) Spectra at the second row (R2) for the angle $\psi_{1}$ (experiment) 
and $\psi = +0.11^{\circ}$ (calculation).
(b) Spectra at the fourth row (R4) for the angle $\psi_{1}$ (experiment) 
and $\psi = +0.11^{\circ}$ (calculation).
(c) Spectra at the second row (R2) for the angle $\psi_{2}$ (experiment) 
and $\psi = -0.02^{\circ}$ (calculation).
The spectra at the top, middle, and bottom are 
those at the left (C1), middle (C2), and right (C3) columns, respectively.
The circles, diamonds, and triangles represent the measured spectra.
The error bars indicate standard errors.
The shaded areas represent the calculated spectra.
Each spectrum is normalized so that maximum value is equal to unity.
}
\label{fig:each}
\end{figure*}

Figure \ref{fig:each} shows the examples of the spectra 
measured at each detector position.
Figure \ref{fig:each} (a) and (b) show the spectra 
measured at the second row (R2) and the fourth row (R4) 
for the angle $\psi_{1}$.
Figure \ref{fig:each} (c) shows the spectra measured at the second row (R2) 
for the different angle $\psi_{2}$.
The sidebands of the sinc-squared function are not clearly observed 
due to the signal-to-noise ratio in the present experiment.
While some of the peak frequency detunings are consistent with zero, 
others are not zero.
In Fig. \ref{fig:each} (a) and (b) 
the spectral peaks shift to the lower frequency detuning region 
if the detector position is changed in the $\theta$ negative direction from C1 to C3.
In Fig. \ref{fig:each} (c) the shift of the spectral peak is not clear.
The FWHMs of the measured spectra are roughly 2--5~GHz.
They are broader than that of the expected sinc-squared spectra (1~GHz).

\begin{figure*}[t]
\centering
\includegraphics[width=18cm,keepaspectratio,pagebox=artbox]{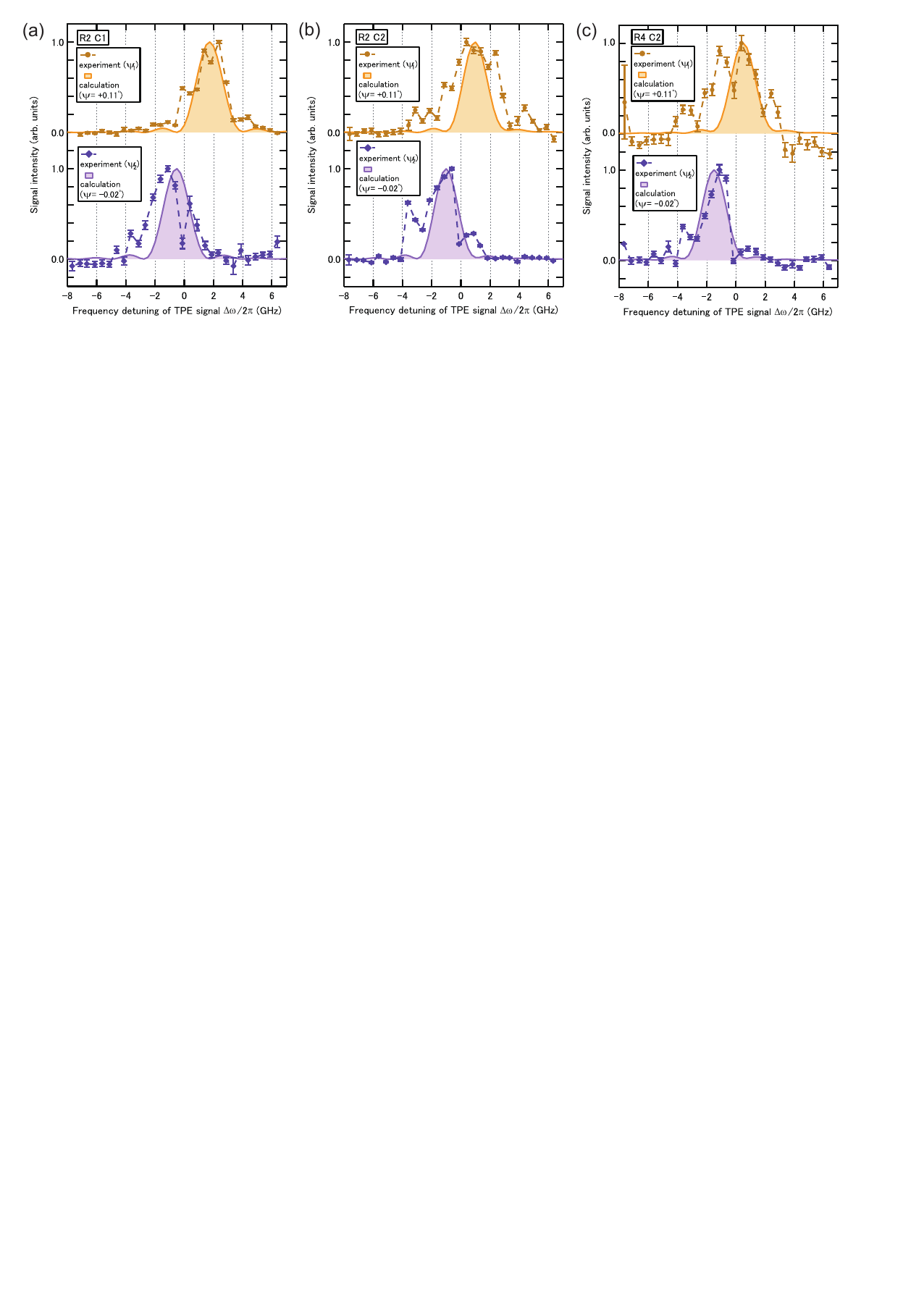}
\caption{(Color online) 
Examples of measured and calculated spectra at each detector position 
for the different angles formed by the counterpropagating pump pulses.
The upper spectra are those for the angle $\psi_{1}$ (experiment) 
and $\psi = +0.11^{\circ}$ (calculation).
The lower spectra are those for the angle $\psi_{2}$ (experiment) 
and $\psi = -0.02^{\circ}$ (calculation).
(a) Spectra at the detector position R2C1.
(b) Spectra at R2C2.
(c) Spectra at R4C2.
Other than the lower spectra of (c) are the same as those in Fig. \ref{fig:each}.
The circles and diamonds are the measured spectra.
The error bars indicate standard errors.
The shaded areas represent the calculated spectra.
Each spectrum is normalized so that maximum value is equal to unity.
}
\label{fig:each_psi}
\end{figure*}

Figure \ref{fig:each_psi} shows 
the examples of the measured spectra of each detector position
for the different angles $\psi_{1}$ and $\psi_{2}$.
The upper and lower spectra are measured for $\psi_{1}$ and $\psi_{2}$, 
respectively.
Other than the lower spectra of Fig. \ref{fig:each_psi} 
(c) are the same as those in Fig. \ref{fig:each}.
The spectra shift to the lower frequency detuning region 
if the angle $\psi$ is changed from $\psi_{1}$ to $\psi_{2}$.
The shifts of the spectral peak are roughly 2~GHz in Fig. \ref{fig:each_psi} (a) and (b) 
and roughly 1~GHz in Fig. \ref{fig:each_psi} (c).

\subsection{Coherent-transient measurement of TPE signal}
\label{subsec:result-timing}

\begin{figure}[t]
\centering
\includegraphics[width=8.5cm,keepaspectratio,pagebox=artbox]{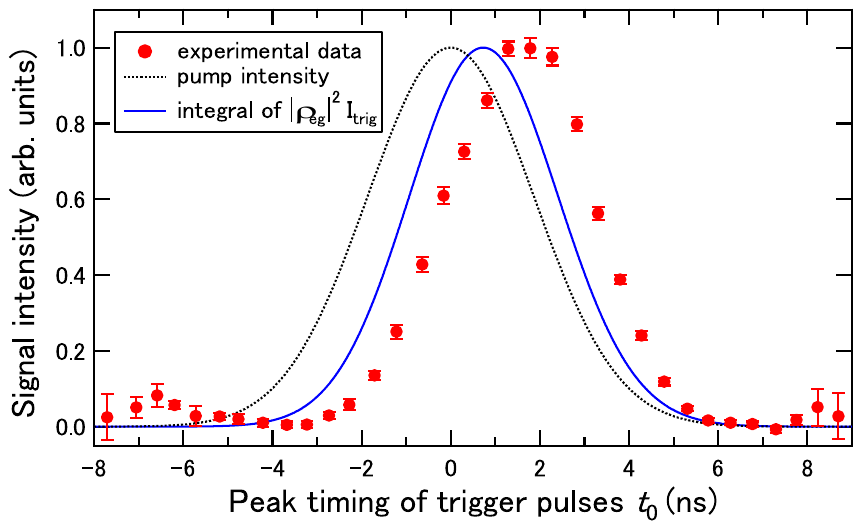}
\caption{(Color online) 
Coherent-transient response 
of the TPE signal energy as a function of 
the peak time of the trigger pulses $t_{0}$.
The red circles represent the experimental data.
The error bars indicate standard errors.
The black dotted line is the intensity of the pump pulses.
The blue solid line is the calculation of 
$\int |\rho_{eg}(t)|^{2} I_{\mathrm{trig}} (t, t_{0}) dt$.
The data and the calculations are normalized 
so that their maximum values are equal to unity.
}
\label{fig:timing}
\end{figure}

We investigate the coherent-transient response of the TPE signal pulse energy 
depending on the mutual timing between the pump and trigger pulses.
The frequency detuning of the TPE signal is fixed at $\Delta \omega = 0$ 
in this measurement.
The detector position is fixed 
and a lens is put between the target and the detector 
to collect the TPE signals as much as possible.
The red circles in Figure \ref{fig:timing} show 
the measured response.
The horizontal axis $t_{0}$ is the peak timing of the trigger pulse 
with respect to that of the pump pulses.
The positive $t_{0}$ indicates that 
the trigger pulse arrives at the target after the pump pulses.
The time $t_{0}$ is binned in 0.5~ns.

The development of the TPE signal intensity shows delayed behavior 
with respect to the pump pulses, 
whose intensities are represented by the black dotted line.
At first the TPE signal intensity monotonically increases.
It has a peak when the trigger pulse is injected roughly 2~ns after the pump pulses.
Then the TPE signal intensity monotonically decreases.
The time width (FWHM) at which the signal pulse is generated is roughly 4~ns.
The small structures at $t_{0} = - 6.5$ and +8.5~ns are not the signal 
but are due to the fluctuation of the background stray photons.

\section{DISCUSSION}
\label{sec:discussion}

\subsection{Spectra of TPE signal}
\label{subsec:discussion-frequency}

\subsubsection{Model}
\label{subsubsec:discussion-frequency-model}

\begin{figure*}[t]
\centering
\includegraphics[width=15cm,keepaspectratio,pagebox=artbox]{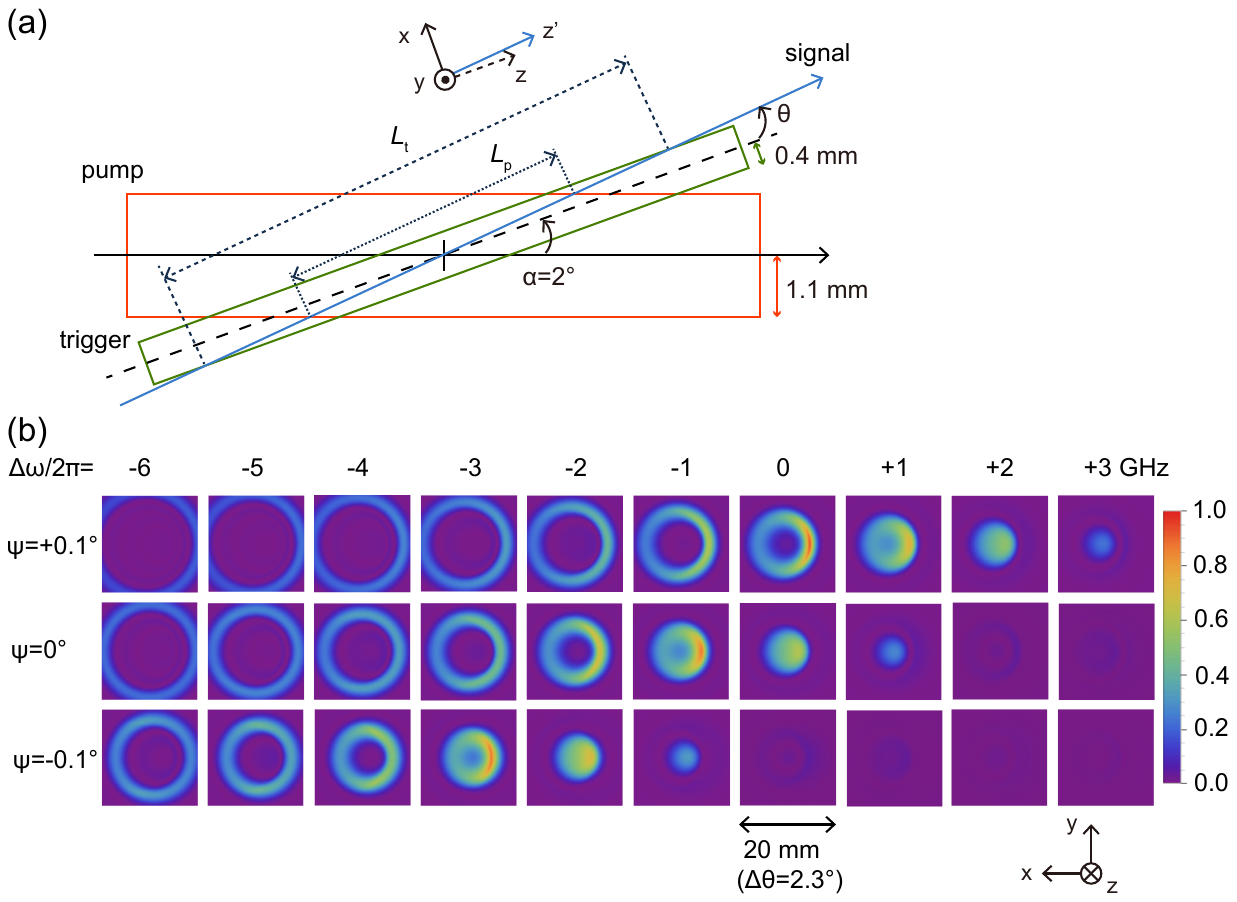}
\caption{(Color online) 
(a) Top view of geometric configuration 
of the pump, trigger, and TPE signal pulses of our model.
The pump and trigger pulses 
are depicted as the red and green rectangles.
The coordinate system is the same as that in Fig. \ref{fig:setup} (a).
The $z$-axis is defined as the central axis of the trigger pulses.
The coordinate origin is at the center of the target.
The blue line represents the path of generated TPE signal.
It passes through the center of the target in our model.
The position $z'$ is defined along the propagation direction of the TPE signal.
The origin of the $z'$-axis is also at the center of the target.
The actual angles $\alpha$ and $\theta$ are very small, 
but they are exaggerated.
(b) Spatial distributions of calculated signal intensity at each position 
for various pairs of $(\Delta \omega,\psi)$.
The calculated region is $-10~\mathrm{mm} \leq x \leq +10~\mathrm{mm}$ 
and $- 10~\mathrm{mm} \leq y \leq +10~\mathrm{mm}$, 
which corresponds to $\Delta \theta = 2.3^{\circ}$.
The spectrum intensity is normalized 
and the color bar does not indicate the absolute intensity.
}
\label{fig:spatial}
\end{figure*}

We discuss the spectra of the TPE signal by using a simple model.
Usually theoretical studies on this process have been based on 
the one-dimensional coupling equations 
in which the sum of $\bm{k}$-vectors of the two counterpropagating pump fields is zero 
and the signal field propagates in the exactly opposite direction to that of the trigger field 
\cite{two-photon-NDFWM-1981}.
In our model we take into account the slight angle 
between two counterpropagating pump fields 
and that between the trigger and signal fields. 
Figure \ref{fig:spatial} (a) shows 
the geometric configurations of the pump, trigger, and TPE signal pulses of our model.
We assume that the TPE signal pulses pass through the center of the target 
and propagate along the $z'$ axis which forms the angle of $\theta$ with the $z$ axis.
The direction of the TPE signal pulses $\theta$ 
has one-to-one correspondence with the position of the MCT detector 
under this assumption.
We also assume that 
the pump and trigger beams are the flat beams 
and the TPE signal photons are generated within the overlapping region of them.
We neglect decrease of the pump and trigger fields 
and the dispersion of p-H$_{2}$ gas in the midinfrared region.

While our model is a quasi-three-dimensional one, 
the equation we use is in one-dimension for simplicity.
The coupling equation 
for the envelope of the electric field of the TPE signal $E_{\mathrm{sig}}$ 
as a function of position $\bm{z}' = z' \hat{\bm{z}}'$ 
is given as follows:
\begin{equation}
\label{eq:main}
\frac{d E_{\mathrm{sig}}}{d z'} = i \kappa \rho_{eg} E_{\mathrm{trig}}^{*} e^{i \Delta \bm{k} \cdot \bm{z}'} .
\end{equation}
Here $\kappa$ is a coupling constant between the trigger and TPE signal fields, 
$\rho_{eg}$ is the off-diagonal element of the density matrix, 
and $E_{\mathrm{trig}}$ is the envelope of the electric field of the trigger field.
$\Delta \bm{k}$ is the $\bm{k}$-vector mismatch: 
\begin{equation}
\Delta \bm{k} = ( \bm{k}_{\mathrm{pump1}} + \bm{k}_{\mathrm{pump2}}) - ( \bm{k}_{\mathrm{trig}} + \bm{k}_{\mathrm{sig}}) .
\end{equation}
Here $\bm{k}_{i} (i=\mathrm{pump1,pump2,trig,sig})$ are 
the $\bm{k}$-vectors of the pump1, 2, trigger, and TPE signal pulses, respectively.
As can be seen from Eq. (\ref{eq:main}), 
what contributes to the development of the TPE signal field 
is not $\Delta \bm{k}$ itself 
but the component parallel to the $z'$ axis of $\Delta {\bm{k}}$.
We represent it as $\Delta k_{\parallel}$.
Solving Eq. (\ref{eq:main}) under the condition 
that $\kappa$, $\rho_{eg}$, and $E_{\mathrm{trig}}$ are constant with respect to $z'$, 
the output TPE signal field is as follows:
\begin{equation}
\label{eq:numerical-signal}
|E_{\mathrm{sig}}|^{2} = | \kappa |^{2} | \rho_{eg} |^{2} |E_{\mathrm{trig}}|^{2} {L_{\mathrm{eff}}}^{2} \mathrm{sinc}^{2} \biggl( \frac{\Delta k_{\parallel} L_{\mathrm{eff}}}{2} \biggr) .
\end{equation}
$L_{\mathrm{eff}}$ is the effective length where the TPE signal is generated.
The condition of $\kappa$, $\rho_{eg}$, and $E_{\mathrm{trig}}$ 
holds in the present experiment.
$\kappa$ is independent of $z'$.
$\rho_{eg}$ is regarded as spatially uniform %constant with respect to $z'$ 
because the electric fields of the pump pulses are 
regarded as spatially uniform in the target cell.
The lengths of the pump pulses (FWHMs) are 1.3~m, 
that are much longer than the target cell 15~cm.
$E_{\mathrm{trig}}$ is also regarded as spatially uniform in the target cell 
and the depletion of $E_{\mathrm{trig}}$ is neglected in our model.

We estimate $\Delta k_{\parallel}$ of our model.
The unit vector $\hat{\bm{z}}'$ 
is represented in spherical coordinates as 
$\hat{\bm{z}}' = (\sin \theta \cos \phi, \sin \theta \sin \phi, \cos \theta)$.
We assume that the two pump pulses are completely superposed in the horizontal plane.
Then the $\bm{k}$-vectors of them are represented as 
$\bm{k}_{\mathrm{pump1}} = k_{\mathrm{pump}} (\sin \alpha, 0, - \cos \alpha)$ 
and $\bm{k}_{\mathrm{pump2}} = k_{\mathrm{pump}} (- \sin (\alpha - \psi) , 0, \cos (\alpha - \psi))$.
The $\bm{k}$-vectors of the trigger pulses and TPE signal pulses are represented as 
$\bm{k}_{\mathrm{trig}} = (0,0,-k_{\mathrm{trig}})$ 
and $\bm{k}_{\mathrm{sig}} = k_{\mathrm{sig}} \hat{\bm{z}}'$.
Therefore $\Delta k_{\parallel}$ is 
\begin{align}
\label{eq:deltak-parallel-angle}
\Delta k_{\parallel} = & \Delta \bm{k} \cdot {\hat{\bm{z}}}' \notag \\
= & 2 k_{\mathrm{pump}} \sin {\frac{\psi}{2}} \biggl\{ \cos \bigg( \alpha - \frac{\psi}{2} \bigg) \sin \theta \cos \phi \notag \\
& + \sin \bigg( \alpha - \frac{\psi}{2} \biggr) \cos \theta \biggr\} + k_{\mathrm{trig}} \cos \theta - k_{\mathrm{sig}} .
\end{align}

$L_{\mathrm{eff}}$ is determined as 
the length of the straight line passing through the center of the target 
inside the overlapping area of the pump and trigger pulses.
The pump and trigger beams are regarded as the cylinders 
with the radii of 1.1~mm and 0.4~mm.
$L_{\mathrm{eff}}(\theta,\phi)$ is 
the shorter of the two lengths $L_{\mathrm{p}}$ and $L_{\mathrm{t}}$.
$L_{\mathrm{p}}$ is twice the distance from the center of the target 
to the intersection point of $z'$ and the cylindrical surface of the pump beam, 
and $L_{\mathrm{t}}$ is that of the trigger beam.
The lengths $L_{\mathrm{p}}$ and $L_{\mathrm{t}}$ 
are shown in Fig. \ref{fig:spatial} (a).
$L_{\mathrm{p}}$ is adopted as $L_{\mathrm{eff}}$ in this case.
Because the angle $\psi$ is very small, 
we assume that 
the cylindrical region of the pump beams are determined only by the pump1 beam 
and it is fixed even if the angle $\psi$ is changed.

TPE signal intensity with each direction of $(\theta, \phi)$ is 
calculated as a function of $(\Delta \omega, \psi)$.
The middle row in Fig. \ref{fig:spatial} (b) shows 
the calculation of the spatial distributions of the TPE signal pulses 
at the position of the detector for the case of $\psi=0^{\circ}$.
Center of each figure corresponds to the position of (0, 0, 500)~mm.
The TPE signal intensities tend to be higher 
in the right sides ($x < 0$) than in the left sides ($x > 0$).
It is because $L_{\mathrm{eff}}$ for $\cos \phi < 0$ ($x < 0$) 
is longer than that for $\cos \phi > 0$ ($x > 0$).
Furthermore, the TPE signal pulses are generated more 
in the negative frequency detuning region $\Delta \omega < 0$.
It is derived from the phase matching condition 
$\Delta k_{\parallel} (\psi = 0^{\circ} ) = k_{\mathrm{trig}} \cos \theta - k_{\mathrm{sig}} = \{ ( \omega_{\mathrm{pump}} - \Delta \omega ) \cos \theta - ( \omega_{\mathrm{pump}} + \Delta \omega ) \} / c = 0$.
This condition means that 
$\Delta \omega = - \frac{1 - \cos \theta}{1 + \cos \theta} \omega_{\mathrm{pump}} < 0$ 
and that $\omega_{\mathrm{sig}}$ is lower than $\omega_{\mathrm{trig}}$.

The top and bottom rows in Fig. \ref{fig:spatial} (b) show 
the spatial distributions of the calculated TPE signal pulses 
for the case of $\psi= + 0.1^{\circ}$ and $- 0.1^{\circ}$, respectively.
The spectrum shifts to the higher frequency detuning region for $\psi > 0^{\circ}$ 
and it shifts to the lower frequency detuning region for $\psi < 0^{\circ}$.
It is also derived from the phase matching condition $\Delta k_{\parallel} = 0$.

\subsubsection{Comparison of measured and calculated spectra}
\label{discussion-frequency-comparison}

We compare the experimental results and the numerical calculation.
The first thing to do is 
to determine which region in calculation corresponds to the measured region.
The size of the measured region is 6~mm horizontal and 8~mm vertical, respectively.
We divide the calculated region into 2~mm $\times$ 2~mm 
with reference to the position of (0, 0, 500)~mm 
and estimate signal intensity in each region.
Numerically calculated spectrum for each position is obtained 
by spatially integrating the right hand side of Eq. (\ref{eq:numerical-signal}) 
over 2~mm $\times$ 2~mm region at each TPE signal frequency.
Assuming that the measured region corresponds to 
the calculated region with the high integrated signal intensity, 
the measured region corresponds to 
the calculated region of $ -6~\mathrm{mm} \leq x \leq 0~\mathrm{mm}$ 
and $ -4~\mathrm{mm} \leq y \leq +4~\mathrm{mm}$.
Unlike the measured data, however, 
the calculated TPE signal intensity does not decrease drastically in the outer region.

The second thing to do is 
to determine the angle $\psi$ for the measured data.
The angle $\psi$ is determined 
so that the center frequency detunings of calculated and measured spectra 
at each detector position coincide.
The center frequency detuning $\Delta \omega_{0,\mathrm{calc}}$ 
of the calculated spectrum of each region is obtained by the fit of 
$A {\mathrm{sinc}}^{2} (B ( \Delta \omega - \Delta \omega_{0} ) )$
with $A$, $B$, and $\Delta \omega_{0}$ as the free parameters.
Because the angle $\theta$ is very small, 
the right hand side of Eq. (\ref{eq:numerical-signal}) is reduced to this function 
and integrated spectrum for each region is very close to this integrand.
We obtain each center frequency detuning 
$\Delta \omega_{0,\mathrm{calc}} ( n , \psi )$ 
for the eleven numbered regions $n$ and various angles $\psi$.
The center frequency detuning of each measured spectrum 
$\Delta \omega_{0,\mathrm{meas}} ( n , \psi_{1(2)} )$ is also obtained 
by the fit of the same function.
Then we calculate the sum of squares of differences between 
$\Delta \omega_{0,\mathrm{calc}} ( n , \psi )$ 
and $\Delta \omega_{0,\mathrm{meas}} ( n , \psi_{1(2)} )$ for each region 
$S(\psi_{1(2)},\psi) = \textstyle\sum_{n=1}^{11} [\Delta \omega_{0,\mathrm{meas}} ( n , \psi_{1(2)} ) - \Delta \omega_{0,\mathrm{calc}} ( n,\psi ) ]^{2}$.
The angle $\psi$ that minimizes $S(\psi_{1(2)},\psi)$ 
is determined by the fit of the quadratic function.
The fitting result gives angles of $\psi = +0.11^{\circ} \pm 0.01^{\circ}$ 
for the data of $\psi_{1}$ 
and $\psi = - 0.02^{\circ} \pm 0.01^{\circ}$ for those of $\psi_{2}$, respectively.
The error bars are scaled so that the reduced $\chi^{2}$s are equal to unity.
The change of angle is 
$\Delta \psi _\mathrm{calc} = -0.13^{\circ} \pm 0.01^{\circ}$.
It is consistent with that measured by the beam profiler, 
$\Delta \psi _\mathrm{meas} = - 0.13^{\circ}$.

We compare spectrum at each position.
The shaded areas in Fig. \ref{fig:each} are 
calculated spectra of each position and each pump angle $\psi$.
The calculated spectra are very close to sinc-squared functions as mentioned above.
While some of the peak frequency detunings are consistent with zero, 
others are not zero as well as the measured spectra.
For the case of $\psi_{1}$ shown in Fig. \ref{fig:each} (a) and (b), 
the shifts of the measured spectra to the lower frequency detuning region 
for the shift of the column to negative $\theta$ direction 
are also reproduced in calculated spectra.
For the case of $\psi_{2}$ shown in Fig. \ref{fig:each} (c), 
the calculated spectra shift to the lower frequency detuning region, 
but the shift is smaller than those in Fig. \ref{fig:each} (a) and (b).
It is consistent with the result 
that the shift of the measured spectra in Fig. \ref{fig:each} (c) 
is not so clear as those in Fig. \ref{fig:each} (a) and (b).
The widths of both measured and calculated spectra are broader than 
that of the expected sinc-squared function (1~GHz).
Furthermore the widths of measured spectra are a little broader than those calculated.
It is because the pressure broadening of p-H$_{2}$ gas 
and linewidths of laser pulses are not considered in our model.

We compare the shifts of the measured spectra at each position 
for the change of the pump angle $\psi$.
The shaded areas in Fig. \ref{fig:each_psi} are calculated spectra.
The shifts to the lower frequency detuning region are reproduced in calculated spectra.

We compare the sums of TPE signal intensity at each position shown in Fig. \ref{fig:total}.
The red and blue shaded areas represent the calculated spectra 
at $\psi = +0.11^{\circ}$ and $\psi = -0.02^{\circ}$, respectively.
The shapes of the calculated spectra are not sinc-squared functions 
as well as those of the measured spectra.
The asymmetry and the tail in the lower frequency detuning region 
of the spectrum for $\psi_{1}$ are reproduced 
in the calculated spectrum for $\psi = +0.11^{\circ}$.
The peak frequency detunings of the calculated spectra are also not zero 
as well as those of the measured spectra.
The measured shift of the spectra by changing the angle $\psi$ 
is reproduced in the calculated spectra.
The peak frequency detuning for $\psi = -0.02^{\circ}$ 
is roughly 2~GHz lower than that for $\psi = +0.11^{\circ}$.
Even if the pump beams completely counterpropagate to each other, 
the TPE signal spectrum does not have a peak at $\Delta \omega = 0$, 
but it shifts by roughly 1~GHz to the lower frequency detuning side.
The magnitude relation between the spectra for $\psi_{1}$ and $\psi_{2}$ 
is reproduced in the calculation.
Regarding the widths of the calculated spectra, 
the FWHMs are several GHz, 
that are also different from the expected sinc-squared spectra.
The reason of the broader widths is 
that the measured spectra are the sum of the TPE signals 
from the various directions that have different peak frequency detunings 
$\Delta \omega_{0,\mathrm{meas}} ( n , \psi_{1(2)} )$.

The total TPE signal spectrum in Fig. \ref{fig:total} is also different from 
the spectra as a function of the two-photon detuning of the pump photons $\delta$ 
shown in Fig. \ref{fig:pump}.
The spectrum of the pump detuning has a peak at the resonance $\delta = 0$ 
and a width of roughly 0.5~GHz.
It corresponds to observation of a two-photon resonance.
The TPE signal spectrum is not simply observation of a resonance, 
but it is determined by phase matching condition, geometry, and so on.

\subsubsection{Validity of our model}

The TPE signal spectra depend on the directions of the TPE signal photons 
and the angle formed by the counterpropagationg photons.
Our simple model can reproduce these dependences and explain the measured spectra.
Our model is a quasi-three-dimensional one, 
but we use the one-dimensional equation.
This approximation holds because 
only the TPE signal pulses passing through the center of the target 
are taken into account.
However those not passing through the center of the target also exist 
in the actual experiment 
and the TPE signal pulses passing through the various paths interfere.
The fact that our model reproduces the measured spectra 
indicates that the effect of this interference is small.
Other approximations are also used in our model.
The assumption that the pump and trigger beams are the flat beams is reasonable 
because those beams are not tightly focused.
The pulse energies of the pump and trigger fields are assumed to be constant.
They are scarcely affected by the interaction between four waves 
because the pulse energy of the generated TPE signal is much smaller than 
those of the pump and trigger pulses.
The dispersion of p-H$_{2}$ gas in the midinfrared region 
can be neglected 
because the dominant factors of the $\bm{k}$-vector mismatch 
are the difference of the wavelengths 
and the angular relationship between the four waves rather than the dispersion.

\subsection{Coherent-transient measurement of TPE signal}
\label{subsec:discussion-time}

The measured coherent-transient response of 
TPE signal depending on the incident time of the trigger pulse 
is delayed with respect to the pump pulses.
The TPE signal is not just the temporal overlap of the pump and the trigger pulses, 
but it would reflect the time evolution of coherence among p-H$_{2}$ 
as mentioned in Sec. \ref{sec:introduction}.

From Eq. (\ref{eq:numerical-signal}), 
the pulse energy of the TPE signal 
is supposed to be proportional to the integration of 
$\int_{- \infty}^{\infty} |\rho_{eg} (t) |^{2} I_{\mathrm{trig}} ( t , t_{0} ) dt$, 
where $I_{\mathrm{trig}}$ is the intensity of the trigger pulses.
This integral is a function of $t_{0}$.
The trigger pulses work as a probe of time evolution of coherence in the present experiment.
We estimate the coherence $\rho_{eg}$ 
by using the optical Bloch equation:
\begin{equation}
\label{eq:Bloch}
\frac{\partial \rho_{ge}}{\partial t} = i (\Omega_{gg} - \Omega_{ee} - \delta) \rho_{ge} + i \Omega_{ge} (\rho_{ee} - \rho_{gg}) - \gamma_{2} \rho_{ge}. 
\end{equation}
Here $\Omega_{gg}$ and $\Omega_{ee}$ are 
the AC Stark shifts of the ground and excited states, respectively.
$\Omega_{ge}$ is the two-photon Rabi frequency, 
and $\gamma_{2}$ is the transverse relaxation rate.
$\gamma_{2}$ is roughly $2 \pi \times 200$~MHz in the present experiment, 
which is estimated by the pressure broadening \cite{pH2-Raman-1986}.
It is considered unnecessary to use the Maxwell-Bloch equations 
because the electric fields of the pump pulses are 
regarded as spatially uniform in the target cell.
The change of the electric fields of the pump pulses are very small 
because the lengths of the pump pulses 
are much longer than that of the target cell as mentioned above.
$\rho_{ee}, | \rho_{ge} | \ll \rho_{gg}$ and $\delta = 0$ 
in the present experiment.
$\Omega_{gg}$ and $\Omega_{ee}$ are much smaller than $\gamma_{2}$.
In this situation Eq. (\ref{eq:Bloch}) is approximated to 
\begin{equation}
\label{eq:Bloch-approximate}
\frac{\partial \rho_{ge}}{\partial t} = - i \Omega_{ge} - \gamma_{2} \rho_{ge} . 
\end{equation}

We assume that the time profiles of the pump fields are described as 
$I_{\mathrm{pump1(2)}} ( t ) = I_{\mathrm{0p}} \exp{\Big[ - 4 \ln 2 \big( \frac{t}{t_{\mathrm{p}}} \big)^{2} \Big]}$.
Here $t_{\mathrm{p}}$ is the duration (FWHM) of the pump pulses.
Because the dominant excitation process related to the TPE signal that we measure is 
the process in which one pump1 photon and one pump2 photon are absorbed, 
$\Omega_{ge}$ is approximately proportional to 
$\sqrt{I_{\mathrm{pump1}} I_{\mathrm{pump2}}}$.
Therefore it is represented as 
$\Omega_{ge} ( t ) = \Omega_{ge} ( 0 ) \exp{\Big[ - 4 \ln 2 \big( \frac{t}{t_{\mathrm{p}}} \big)^{2} \Big]}$.
With the boundary condition of $\rho_{ge} (t = - \infty)=0$, 
the solution of Eq. (\ref{eq:Bloch-approximate}) is 
\begin{align}
\label{eq:rho}
\rho_{ge} (t) = & - \frac{i}{4} \sqrt{\frac{\pi}{\ln 2}} \Omega_{ge} ( 0 ) t_{\mathrm{p}} \exp{ \Big[ {\Big( \frac{\gamma_{2} t_{\mathrm{p}}}{4 \sqrt{\ln 2}} \Big)^{2} - \gamma_{2} t} \Big] } \notag \\
& \Big\{ 1+ \mathrm{erf}{\Big( 2 \sqrt{\ln 2} \frac{t}{t_{\mathrm{p}}} - \frac{\gamma_{2} t_{\mathrm{p}}}{4 \sqrt{\ln 2}} \Big)} \Big\} .
\end{align}
The time dependence of the coherence 
normalized by the Rabi frequency $\Omega_{ge}(0)$ 
is independent of the pump pulse intensities.
It holds in the limit of the weak pump pulse intensities 
where $\rho_{ee}$ and $| \rho_{ge} |$ are much smaller than $\rho_{gg}$.

We also assume that the time profile of the trigger pulse intensity is represented as 
$I_{\mathrm{trig}} ( t, t_{0} ) = I_{\mathrm{0t}} \exp{\Big[ - 4 \ln{2 \big( \frac{t - t_{0} }{t_{\mathrm{t}}} \big)^{2}} \Big]}$ 
with the duration (FWHM) of the trigger pulse $t_{\mathrm{t}}$.
The blue solid line in Fig. \ref{fig:timing} is the numerical integration of 
$\int_{- \infty}^{\infty} |\rho_{eg} (t) |^{2} I_{\mathrm{trig}} (t , t_{0}) dt$.
The measured result that the TPE signal intensity is high 
if the trigger pulses are delayed with respect to the pump pulses 
is reproduced in our model.
This delayed behavior can be explained as follows.
The two-photon Rabi frequency is 
$\Omega_{ge} / 2 \pi \sim \mathcal{O} (1)$~MHz in the present experiment.
It indicates that 
the period of Rabi oscillation is much longer than the duration of the pump pulses.
Coherence develops late compared with the pump fields, 
it has a maximum after the peak of the pump pulses, 
and then it decays due to the decrease of the pump field intensities and decoherence.
This delayed behavior was also reported in our previous study 
using the Raman process \cite{external-trigger-2015}.

However, 
the calculational result is roughly 1~ns earlier than the experimental result.
The discrepancy would be explained by the combination of the various effects, 
for example, the actual $\gamma_{2}$, 
the actual pulse durations of the pump and trigger pulses, 
and the deviations from the gaussian pulses.

\section{Conclusions}
\label{sec:conclusion}

We have studied of NDFWM using two-photon-allowed vibrational transition of p-H$_{2}$.
We have observed the coherently amplified TPE signal 
which is generated as a result of the NDFWM process 
and measured dependences on the frequency and incident time of the trigger pulse.

The measured spectra are different from 
the solution of the usual one-dimensional coupling equations.
They vary depending on the direction of the signal photon 
and the angle formed by the pump photons, 
that have not been considered in the previous studies.
The measured geometry-dependence can be explained by a simple model 
using phase matching condition and simplified geometry, 
and neglecting beam intensity profile, depletion of incident fileds, 
and dispersion of p-H$_{2}$ in midinfrared region.
The present study suggests a possible application 
as an optical narrow bandpass retroreflector 
whose reflection frequency is tunable 
by the angle between the counterpropagating pump photons.

The configuration using counterpropagating pump fields 
also has a possible application 
for study of coherent amplification 
of rare emission of plural particles involving neutrinos \cite{PTEP-2012}.
Several properties of neutrinos 
such as absolute masses, mass type (Dirac or Majorana), and $CP$-violating phases 
can be determined from the emission spectrum.
Applying the result obtained in the present paper to this process, 
it can be predicted that careful attention is required 
for discussion about shape of the obtained signal spectrum.
The alignment of the pump fields and the direction of the signal field 
cause systematic errors of the signal spectrum.

The measured dependence of the TPE signal on the incident time of the trigger pulse 
shows roughly 2~ns delayed behavior with respect to the pump pulses.
The measured coherent-transient response 
can be qualitatively explained 
by another simple model using the optical Bloch equation.
The trigger field has a role as a probe of the time evolution of the coherence 
generated by the pump fields.
It is necessary to slightly delay the incident time of the trigger pulse 
with respect to those of the pump pulses 
in order to increase the output intensity of the bandpass retroreflector.

\begin{acknowledgments}
We thank Professor J. Tang for lending the seeding laser for the pulsed Nd:YAG laser.
This work was supported by JSPS KAKENHI 
(Grants No. JP15H02093, No. JP15H03660, No. 15K13486, No. JP16J10938, 
No. JP17H02895, No. JP1702896, No. JP17K14292, No. JP17K14363, and No. JP17K18779) 
and JST, PRESTO Grant No. JPMJPR16P3.
\end{acknowledgments}

%\bibliography{BibTPEwavelength2}
%\bibliography{BibCounterExternal}
%\providecommand{\noopsort}[1]{}\providecommand{\singleletter}[1]{#1}%
%\begin{thebibliography}{18}%
%\end{thebibliography}%

\providecommand{\noopsort}[1]{}\providecommand{\singleletter}[1]{#1}%

\end{document}